\documentclass[twocolumn,aps,prb,showpacs,superscriptaddress]{revtex4}
\usepackage{mathrsfs,bm,multirow,amsmath,amsfonts,amssymb,array,booktabs,float}
\usepackage{graphicx,times,graphics,color,epsfig}

\begin{document}
\title{Two-Dimensional Photogalvanic Spin-Battery}

\author{Yiqun Xie}
\affiliation{Department of Physics, Shanghai Normal University, 100 Guilin Road, Shanghai 200232, China}
\affiliation{Department of Physics and the Center of Theoretical and Computational Physics, The University of Hong Kong, Pokfulam Road, Hong Kong SAR, China}
\author{Mingyan Chen}
\affiliation{Department of Physics, Shanghai Normal University, 100 Guilin Road, Shanghai 200232, China}
\affiliation{Department of Physics and the Center of Theoretical and Computational Physics, The University of Hong Kong, Pokfulam Road, Hong Kong SAR, China}
\affiliation{Hongzhiwei Technology (Shanghai) Co., Ltd., 1888 Xinjinqiao Road, Shanghai 201206, China}
\author{Zewen Wu}
\affiliation{Department of Physics and the Center of Theoretical and Computational Physics, The University of Hong Kong, Pokfulam Road, Hong Kong SAR, China}
\affiliation{Department of Physics, Beijing Institute of Technology, Beijing 100081, China}
\author{Yibin Hu}
\affiliation{National Laboratory for Infrared Physics, Shanghai Institute of Technical Physics, Chinese Academy of Sciences, Shanghai 200083, China}
\author{Yin Wang}
\email{yinwang@shu.edu.cn}
\affiliation{Department of Physics and the Center of Theoretical and Computational Physics, The University of Hong Kong, Pokfulam Road, Hong Kong SAR, China}
\affiliation{Department of Physics and International Centre for Quantum and Molecular Structures, Shanghai University, 99 Shangda Road, Shanghai 200444, China}
\author{Jian Wang}
\affiliation{Department of Physics and the Center of Theoretical and Computational Physics, The University of Hong Kong, Pokfulam Road, Hong Kong SAR, China}
\author{Hong Guo}
\affiliation{Department of Physics and the Center of Theoretical and Computational Physics, The University of Hong Kong, Pokfulam Road, Hong Kong SAR, China}
\affiliation{Center for the Physics of Materials and Department of Physics, McGill University, Montreal, Quebec H3A 2T8, Canada}

\date{\today}
\begin{abstract}
Pure spin-current is of central importance in spintronics. Here we propose a two-dimensional (2D) spin-battery system that delivers pure spin-current without an accompanying charge-current to the outside world at zero bias. The principle of the spin-battery roots in the photogalvanic effect (PGE), and the system has good operational stability against structural perturbation, photon energy and other materials detail. The device principle is numerically implemented in the 2D material phosphorene as an example, and first principles calculations give excellent qualitative agreement with the PGE physics. The 2D spin-battery is interesting as it is both a device that generates pure spin-currents, also an energy source that harvests photons. Given the versatile operational space, the spin-battery should be experimentally feasible.
\end{abstract}
\maketitle

\emph{NOTE:} This article has been published in Physical Review Applied in a revised form (DOI: 10.1103/PhysRevApplied.10.034005).
\section{Introduction}

Two dimensional (2D) materials are important for applications in logic and photonic devices, solar cells, transparent substrates and most interestingly, wearable electronics\cite{nnano-review}. The thin body in 2D makes them the natural choice for producing flexible structures. The spin physics of 2D materials is central for flexible spintronics that requires less power to operate\cite{flexiable,spintronics,RevSpin, Hu-review}. In flexible applications, self-powered systems - by harvesting photons for example, are greatly helpful\cite{flexiable}. It is the purpose of this work to propose and investigate a 2D photogalvanic spin-battery that generates pure spin-current without an accompanying charge current, for 2D spintronics.

The photogalvanic effect (PGE) is an optoelectronic phenomenon that occurs in materials without a spatial inversion symmetry\cite{PGE0}. A \emph{direct} charge current is generated to flow by PGE with neither external bias voltage nor internal electric field like that inside the p-n junction of photo-cells. PGE is purely a \emph{nonlinear} and \emph{symmetry induced} optical response of materials to light, $j \sim \alpha \textit{\textbf{E}}\textit{\textbf{E}}^*$, where $j$ is the PGE photocurrent, \textit{\textbf{E}} is the electric field of the light, and $\alpha$ the PGE coefficient. Because reversing direction of \textit{\textbf{E}} will reverse the flow of current, $j$ to $-j$, $j$ must vanish unless $\alpha$ also changes its sign which can only occur when the material has no spatial inversion symmetry\cite{PGE0}. A device operating on PGE can deliver a dc electric current to the outside world in close circuit, i.e. liking a battery. Recent experimental demonstrations of PGE include using silicon nanowires\cite{Agarwal} and Si MOSFETs\cite{Si-Mos2,Si-Mos1}, where inversion symmetry was broken by geometry of the device.

PGE has been exploited to generate spin-polarized charge current\cite{PGE1,PGE2,PGE3,PGE3a,PGE4}, typically in materials having a strong spin-orbit coupling (SOC) including transition metal dichalcogenides \cite{PGE5,PGE6} and topological insulators\cite{PGE7,TI2,TI3}. These materials lift spin degeneracy by SOC so that spin-up and -down electrons are photo-excited by left and right circularly polarized light governed by the optical selection rule, to produce spin-polarized photocurrent $I_{ph}=I_{\uparrow} + I_{\downarrow}$, where $I_{\uparrow}, I_{\downarrow}$ are contributions from the spin-up and -down channels. Clearly, accompanying the charge-current $I_{ph}$, there is also a spin-current $I_s=I_{\uparrow} - I_{\downarrow}$, which is nonzero if $I_{\uparrow} \neq I_{\downarrow}$. A most interesting situation is when $I_{\uparrow} = -I_{\downarrow}$, i.e. the spin-up and -down channels flow in exactly opposite directions. When this happens, a \emph{pure} spin-current $I_s$ flows without an accompanying charge-current $I_{ph}$. Pure spin-current is prominent in spintronics\cite{spintronics}, spin caloritronics\cite{spc, Tian}, spin Hall\cite{HE1,opt4}, spin pumping\cite{sp1,sp2,sp3} and spin Seebeck effects\cite{sps1,sps2,sps3,sps4, sierra}, where SOC plays the key role in lifting the spin degeneracy and/or inducing a transverse anomalous velocity.

Can PGE provide a physical principle of a device that generates pure spin-current without the aid of SOC? Indeed, in wide category of materials consisting light elements such as carbon, SOC is negligibly weak. Since there is no SOC, we consider such a device design (hereafter named spin-battery) in the form of a thin semiconductor (S) barrier sandwiched between two ferromagnetic metal (FM) contacts.
The FM/S/FM structure is widely used in tunnel junctions where both spin-up and -down electrons flow from one FM through the semiconductor to the other FM, i.e. a flow of spin-polarized charge-current accompanied by a spin-current\cite{PRB-1,PRB-2,PRB-3,PRB-4}. By PGE, however, we show that the spin-battery can generate a pure $I_s$ to flow out of the system without any charge-current, and in open circuit, a spin resolved chemical potential difference is established on the two sides of the device due to the action of PGE.

\section{Physics of the PGE spin-battery}
\begin{figure}
\centering
\includegraphics[width=0.9\columnwidth]{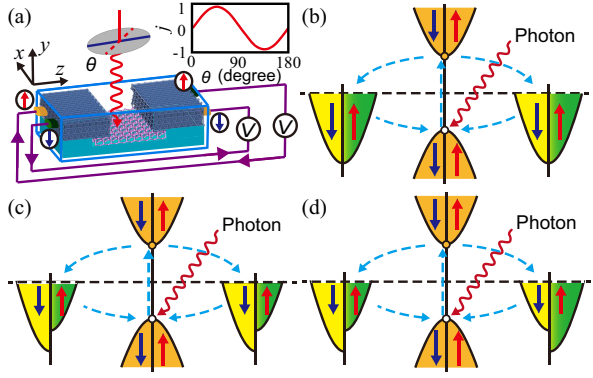}
\caption{(a) Schematic diagram of the spin-battery which is a FM/semiconductor/FM structure under illumination of polarized light, shown in the blue cuboid. The spin-down/up electrons come out from the system, flow through the external circuit, and finally, go back to the spin-battery via the orange electrode. (b) Band diagram of spin-degenerate case. (c) Band diagram for parallel configuration of the two FM in the system. (d) Band diagram for antiparallel configuration of the two FM, this is the spin-battery device. The blue dashed lines with arrows indicate electron transitions.}
\label{fig1}
\end{figure}

The proposed PGE spin-battery is in the form of FM/S/FM in open circuit, and in close circuit it is part of a two-probe structure, (FM-electrode)+(FM/S/FM)+(FM-electrode), shown in Fig.~\ref{fig1}. Assuming there is no spatial inversion symmetry and for simplicity of discussion let's consider the central region to be mirror symmetric across the middle ($C_s$ point group). The semiconductor has an appropriate band gap for photon adsorption. Without losing generality, we implement the idea in two-dimensional (monolayer) black phosphorous called phosphorene as the semiconductor, sandwiched by nickel FM contacts. Phosphorene has attractive electronic and optical properties including significant carrier mobility\cite{zhang}, direct band gap\cite{Ji,GW}, broadband photoresponse\cite{BpOpt0,BpOpt1,BpOpt2,BpOpt3} and spin transmission\cite{small,AM}. We emphasize that the physics of PGE spin-battery is general and the semiconductor is not restricted to phosphorene.

Before discussing the principle of spin-battery, let's outline PGE from the point of view of photo-excitation, using Fig.~\ref{fig1}(b) which describes spin-degenerate case by replacing the FM metal with nonmagnetic metal such as Cu or Au. By absorbing photons, electrons are excited from the valence bands (VB) to the conduction bands (CB), and they flow out of the semiconductor to the left or right. At the same time, the holes left in VB are filled with electrons coming from the two metals. These processes are indicated by the blue dashed lines in Fig.~\ref{fig1}(b). When incident light is polarized in parallel (polarization angle $\theta=90^\circ$) or perpendicular ($\theta=0^\circ,\ 180^\circ$) to the mirror reflection plane, $C_s$ symmetry is maintained and due to this mirror symmetry, excited electrons have exactly the same probability to flow to the left or right, resulting in zero photocurrent. For other polarization angles $\theta$, the coupling of polarized photons and electrons in the semiconductor breaks mirror symmetry, hence photo-excited electrons have different probabilities to flow to the left or right, resulting to a net photogalvanic current $I_{ph}$ at zero bias. This PGE current has a distinct dependence on the polarization angle $\theta$, e.g. varies as $\sin(2\theta)$ for the case of $C_s$ symmetry.

The physical origin of the photocurrent generated by PGE can be understood from the standard phenomenological theory\cite{PGE0}. For a material with Cs symmetry the photocurrent under normal incidence of a linearly polarized light is described by
\begin{equation}\label{j}
j_z=E^2_0\chi_{zzx}\sin(2\theta), \ \ j_x=E^2_0(\chi_++\chi_-\cos(2\theta)).
\end{equation}
Here,  $j_z$ and $j_x$ are the photocurrents along the directions normal and parallel to the mirror reflection plane,  respectively.
For the circularly polarized light,
\begin{equation}\label{j2}
j_z=E^2_0\gamma_{zy}\sin(2\phi), \ \ j_x=E^2_0(\chi_++\chi_-\cos(2\phi)).
\end{equation}
In Eqs.(\ref{j}) and (\ref{j2}), $E^2_0$ is the electric field amplitude of the light, $\chi_{zzx}$, $\chi_+$, $\chi_-$  and $\gamma_{zy}$ are tensors which depend on the photon frequency $\omega$. Therefore, the PGE is essentially a second-order response to the electric field. Within this scenario,  the light illumination contributes to the response coefficients $\chi_{zzx}$, $\chi_+$, $\chi_-$  and $\gamma_{zy}$, and thus generates a PGE photocurrent. Therefore, the behavior of the photocurrent either takes a sine or a cosine dependence on the light polarization, determined by both the light and the symmetry of the system.

Having understood the ordinary PGE, we now discuss cases with spins. If magnetic moments of the two FM in the spin-battery are in parallel configuration (PC) as shown in Fig.~\ref{fig1}(c), photo-excited electrons - regardless of their spin, act just like that in the nonmagnetic case, because the two spin channels are not coupled (no SOC). Consequently a spin-polarized charge-current $I_{ph}$ is generated by PGE when $\theta$ is not $0^\circ,\ 90^\circ,\ 180^\circ$.

Finally, if magnetic moments of the FM are in antiparallel configuration (APC) as shown in Fig.~\ref{fig1}(d), the $C_s$ mirror reflection symmetry of the device for \emph{individual} spin channel is broken by the majority/minority density of states in the FM. Then, both spin-up and -down PGE currents $I_{\uparrow}, I_{\downarrow}$ are generated for \emph{all} angles $\theta$, including at $\theta=0^\circ$, $90^\circ$ and $180^\circ$. Note that even though the spin distribution is different in APC as compared to PC, the \emph{charge distribution} is the same in PC and APC (also nonmagnetic case). Therefore the total \emph{charge-current} $I_{ph}=I_{\uparrow}+I_{\downarrow}$ must behave the same way as that in PC and/or in nonmagnetic case, namely $I_{ph}$ must vanish at $\theta=0^\circ$, $90^\circ$ and $180^\circ$. We therefore arrive at a most interesting conclusion for APC: namely since $I_{\uparrow}, I_{\downarrow}$ are nonzero at $\theta=0^\circ,\ 90^\circ,\ 180^\circ$ by PGE but $I_{ph}$ vanishes at these angles, $I_{\uparrow}$ and $I_{\downarrow}$ must be equal in magnitude but flow in exactly opposite direction. In other words, for a device without inversion symmetry but is mirror symmetric, a nonzero spin-current $I_s$ without an accompanying charge-current $I_{ph}$ can be generated by PGE in APC. This device behaves as a spin battery because in open circuit, a spin resolved chemical potential difference is established on the two sides of the device due to the action of PGE. The device in Fig.~\ref{fig1}(d) is the pure spin-battery which will be the focus of the rest of the paper.

\begin{figure}
\centering
\includegraphics[width=0.5\columnwidth]{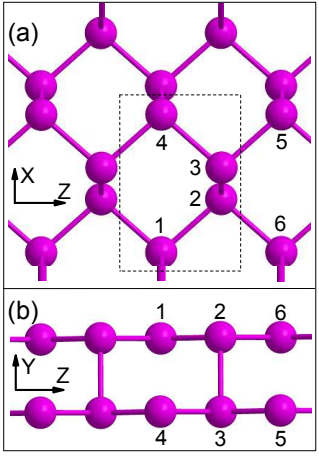}
\caption{(a) The top and (b) side views of the 2D phosphorene. The dashed lines in figure (a) indicate the rectangle primitive cell of the phosphorene which has four atoms.}
\label{fig2}
\end{figure}

Having qualitatively understood the pure spin-battery, we now implement the idea using the 2D phosphorene as the semiconductor. The phosphorene has a puckered structure with four P atoms in primitive cell as shown in Fig.~\ref{fig2}(a). It has a spatial inversion symmetry which belongs to the $D_{2h}$ point group with the inversion center located between the atoms 2 and 3 indicated in Fig.~\ref{fig2}(a). This means that the symmetry is invariant when interchanging the atoms 1, 2 and 6 with atoms 5, 3 and 4, respectively. It is essential to break this intrinsic inversion symmetry to have PGE. This is actually not difficult because the local Schottky electric field formed at the metal/phosphorene contacts breaks the inversion symmetry, as shown experimentally (albeit in a different material)\cite{Agarwal}.

\section{Model and methods}
\begin{figure}
\centering
\includegraphics[width=\columnwidth]{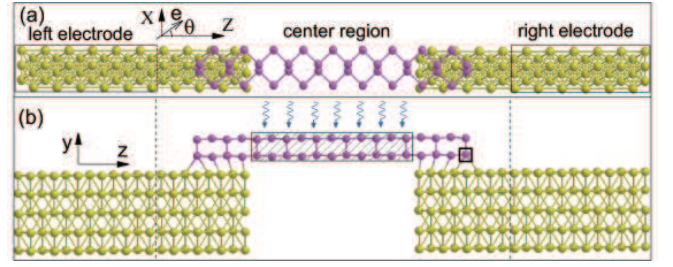}
\caption{(a) The top view and (b) side view of the two-probe device structures for calculating the photocurrent of the phosphorene phototransistor. The yellow spheres denote Ni atoms and others are P atoms. The light (arrows) is shed in the $y$ direction, perpendicularly to the phosphorene in the $x$-$z$ plane. $\theta$ is the polarization angle for the linearly polarized light, and  \textbf{e} is the polarization vectors. The shadows in (b) denote the P atoms illuminated by the light, and the P atom in the rectangle box will be replaced by a As atom (see the main text).}
\label{fig3}
\end{figure}

Hence, we construct a FM/semiconductor/FM structure whose energy diagram is that in Fig.~\ref{fig1}(c,d) where the FM is nickel (Ni) and the semiconductor is phosphorene, as shown in Fig.~\ref{fig3} (a). Phosphorene has a $C_s$ symmetry in which the zigzag direction is perpendicular to the mirror reflection plane. Again, the formation of Ni/phosphorene interface leads to a Schottky potential that breaks the inversion symmetry of phosphorene\cite{Agarwal}. For instance, the electrostatic potential at atom 2 does not equal to that at atom 3, owing to the broken inversion symmetry. Note that the central region of the spin-battery is a composite system of Ni/phophorene/Ni.

The device model consists of three parts including the left- and right-hand  nickel electrodes, and the center region, as shown in Fig.~\ref{fig3}(a). The electrodes are modeled by a five-layer Ni(111) slabs consisting of $2\times3$  atoms in each layer, and the two electrodes extend to $z=\pm \infty$, respectively. The  left- and right-hand
electrodes are the mirror images of each other. In the center region, the phosphorene bridges and partially overlaps the two Ni(111) electrodes, with a vertical Ni-P distance of 2.0 \AA\ determined by the \textsc{vasp} calculation\cite{vasp}. The whole system periodically extends in the $x$ direction  with a periodicity of 4.316 \AA. The calculated lattice constants of the phosphorene are $a$=3.32 \AA, and $b$=4.58 \AA\ which is homogeneously strained by about $-5.7\%$ in order to match the Ni lattice in the $x$ direction. The bandgap  of the phosphorene is  1.03eV calculated with the local density approximation.

The pristine phosphorene has a spatial inversion symmetry and belongs to $D_{2h}$ point group with a mirror reflection plane (the $x$-$y$ plane)  perpendicular to the zigzag ($z$) direction. However,  the contact with the electrodes breaks the $D_{2h}$ symmetry of the pristine phosphorene, which  can be easily seen from the side view of the center region shown in Fig.~\ref{fig3}(b). Even though, the  mirror reflection plane (the $x$-$y$ plane) still remains, as can be observed from Fig.~\ref{fig3}(a). Thus the phototransistor possesses the $C_s$ symmetry,
owing to which we can obtain a spin-polarized photocurrent generated by the PGE.

We have developed a theoretical approach to calculate the photocurrent generated by the PGE based on density functional theory within the nonequilibrium Green's function (NEGF-DFT)
formalism\cite{xie} and implemented it in the first-principles quantum transport package \textsc{nanodcal}\cite{Taylor}. Specifically, for the linearly polarized light  the photocurrent injecting into the lead L can be written as\cite{xie},
\begin{eqnarray}\label{LPGE1}
J^{(ph)}_L&=&\frac{ie}{h}\int\{\cos^2\theta\textrm{Tr}\{\Gamma_L[G^{<(ph)}_{1}+f_L(G^{>(ph)}_{1}-G^{<(ph)}_{1})]\} \nonumber\\
&+&\sin^2\theta\textrm{Tr}\{\Gamma_L[G^{<(ph)}_{2}+f_L(G^{>(ph)}_{2}-G^{<(ph)}_{2})]\} \nonumber\\
&+&2\sin(2\theta)\textrm{Tr}\{\Gamma_L[G^{<(ph)}_{3}+f_L(G^{>(ph)}_{3}-G^{<(ph)}_{3})]\}\}dE, \nonumber\\
\end{eqnarray}
   where
\begin{eqnarray}\label{LPGE}
 G^{>/<(ph)}_{1}&=&\sum_{\alpha,\beta=x,y,z}C_0NG^r_0e_{1\alpha}p^{\dagger}_{\alpha}G^{>/<}_0e_{1\beta}p_{\beta}G^a_0\nonumber\\
 G^{>/<(ph)}_{2}&=&\sum_{\alpha,\beta=x,y,z}C_0NG^r_0e_{2\alpha}p^{\dagger}_{\alpha}G^{>/<}_0e_{2\beta}p_{\beta}G^a_0\nonumber\\
 G^{>/<(ph)}_{3}&=&\sum_{\alpha,\beta=x,y,z}C_0N(G^r_0e_{1\alpha}p^{\dagger}_{\alpha}G^{>/<}_0e_{2\beta}p_{\beta}G^a_0\nonumber\\
 &+&G^r_0e_{2\alpha}p^{\dagger}_{\alpha}G^{>/<}_0e_{1\beta}p_{\beta}G^a_0).\nonumber\\
\end{eqnarray}
For the circularly polarized light, we obtain:
\begin{eqnarray}\label{CPGE1}
J^{(ph)}_L&=&\frac{ie}{h}\int\{\cos^2\phi\textrm{Tr}\{\Gamma_L[G^{<(ph)}_{1}+f_L(G^{>(ph)}_{1}-G^{<(ph)}_{1})]\} \nonumber\\
&+&\sin^2\phi\textrm{Tr}\{\Gamma_L[G^{<(ph)}_{2}+f_L(G^{>(ph)}_{2}-G^{<(ph)}_{2})]\} \nonumber\\
&+&\frac{\sin(2\phi)}{2}\textrm{Tr}\{\Gamma_L[G^{<(ph)}_{3}+f_L(G^{>(ph)}_{3}-G^{<(ph)}_{3})]\}\}dE, \nonumber\\
\end{eqnarray}
where $G^{>/<(ph)}_{1,2}$ is same as that in linear polarized case and
\begin{eqnarray}\label{CPGE}
 G^{>/<(ph)}_{3}&=\pm i&\sum_{\alpha,\beta=x,y,z}C_0N(G^r_0e_{1\alpha}p^{\dagger}_{\alpha}G^{>/<}_0e_{2\beta}p_{\beta}G^a_0\nonumber\\
 &-&G^r_0e_{2\alpha}p^{\dagger}_{\alpha}G^{>/<}_0e_{1\beta}p_{\beta}G^a_0).\nonumber\\
\end{eqnarray}
In Eqs.(\ref{LPGE}) and (\ref{CPGE}), $C_0=(e/m_0)^2 \frac{\hbar\sqrt{\mu_r\epsilon_r}}{2N\omega\epsilon c}I_\omega$ where $m_0$ is the bare electron mass; $I_\omega$ is the photon flux defined as the number of photons per unit time per unit area; $\omega$ is the frequency and $c$ the speed of the light; $\mu_r$ is the relative magnetic susceptibility, $\epsilon_r$ the relative dielectric constant, and $\epsilon$  the dielectric constant. $N$ is the number of photons. In the above expressions, $G^{r/a}_0$ are the retard/advanced Green's functions without photons, and $G^{>/<}_0$ are the greater/lesser Green's function without photons. $p_{x,y,z}$ is the cartesian component of the electron momentum, and $e_{1/2x,y,z}$ is the cartesian component of the unit vector \textbf{e}$_{1/2}$ which characterizes the polarization of the light. For a elliptically polarized light the polarization vector \textbf{e}=$\cos\phi$\textbf{e}$_1$$\pm i\sin \phi$\textbf{e}$_2$, where  ``$\pm$" denotes the right/left handed elliptical light, and $\phi$ determines its helicity. In particular, $\phi=\pm45^o$ corresponds to the circularly polarized light. For a linearly polarized light  \textbf{e}=$\cos\theta$\textbf{e}$_1$+$\sin\theta$\textbf{e}$_2$, where $\theta$ is the angle formed by the polarization direction with respect to the vector $\textbf{e}_1$. We calculate a normalized photocurrent, i.e. the photoresponse function written as $R= \frac{J^{(ph)}_L}{eI_\omega}$,
which still has a dimension of area, i.e., $a_0^2$/phonon where $a_0$ is the Bohr radius.
We note that for laser power of 0.1 to 105 $W/cm^2$ with an illumination area of 1 $\mu m^2$, the calculated photocurrent is from 1 $pA$ to 1 $\mu A$, which can be measured experimentally\cite{acsnano-Wu, BpOpt2}.

We calculate PGE current along the zigzag ($z$) phosphorene direction under linearly or circularly polarized light vertically incident to the phosphorene ($y$ direction), indicated by the wiggly arrows in Fig.~\ref{fig3}(b). The light is polarized in the $x$-$z$ plane, and for linearly polarized light the polarization vector \textbf{e} forms an angle $\theta$ with respect to the zigzag direction. A wide range of photon energies from 1.2 eV to 2.0 eV are investigated. SOC is not considered to reduce computational cost: we verified that including SOC does not qualitatively change any results.
In the NEGF-DFT numerical calculations, double-zeta polarized (DZP) atomic orbital basis was used to expand all the physical quantities; the exchange and correlation were treated at the level of local density approximation; and atomic cores are defined by the standard norm conserving nonlocal pseudopotentials. In the self-consistent calculations and the photocurrent transport calculations of the two-probe device structures (Fig.~\ref{fig3}), $16\times1\times1$ k-points were used. These calculation details were verified to obtain converged results. Unless otherwise specified, collinear spin is used for the NEGF-DFT calculations in which spin polarizations at different places are all along the same direction in either parallel or anti-parallel configuration.

\begin{figure}
\centering
\includegraphics[width=0.9\columnwidth]{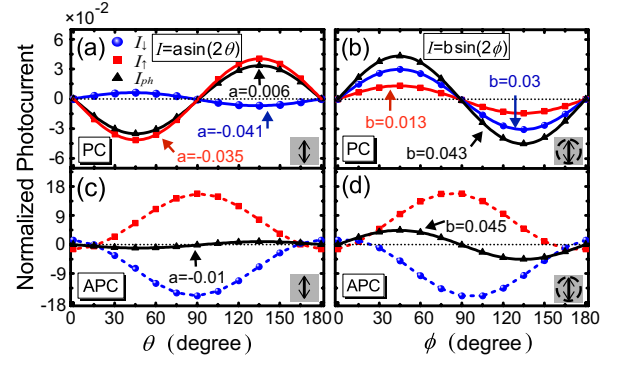}
\caption{The spin-up (I$_{\uparrow}$), spin-down (I$_{\downarrow}$) and total photocurrent (I$_{ph}$) under   linearly- and circularly polarized light for parallel configuration (a,b) in PC and for anti-parallel configuration (c,d) in APC. Symbols denote the calculated results and solid lines denote the fitted ones.}
\label{fig4}
\end{figure}

\section{Results and discussion}
\subsection{PGE spin-battery}
We first consider the case of PC [see Fig.~\ref{fig1}(c)] for which the calculated PGE photocurrent versus the polarization angle $\theta$ for linearly polarized light is presented in Fig.~\ref{fig4}(a). The PGE photocurrent is spin-polarized and both I$_{\uparrow}$ and I$_{\downarrow}$ vary as sin(2$\theta$) and have opposite sign (i.e. opposite flow direction). Specifically, I$_{\uparrow}$ (red square) and I$_{\downarrow}$ (blue sphere) are well fitted as I$_{\uparrow} \approx -0.041\times sin(2\theta)$ (red line) and I$_{\downarrow}\approx 0.006\times sin(2\theta)$ (blue line), respectively. Consequently, the total PGE charge-current, I$_{ph}$=I$_{\uparrow}$ + I$_{\downarrow}$ (dark triangles) also has a sinusoidal shape (black line) which is a characteristic feature of linear photogalvanic effect (LPGE)\cite{PGE0,PGE2}. For circularly polarized light, Fig.~\ref{fig4}(b) shows that the I$_{\uparrow}$ and I$_{\downarrow}$ have the same sign, and both vary in terms of sin(2$\phi$) where $\phi$ is the helicity of the circularly polarized light. The photocurrents are also well fitted with sine function (lines) as shown in Fig.~\ref{fig4}(b). The sine dependence of photocurrent on the helicity of circularly polarized light is the well known circular PGE (CPGE)\cite{PGE2,PGE3,PGE3a,PGE4}. These numerical results are perfectly consistent with the qualitative discussion above: for instance at $\theta=0^\circ$, $90^\circ$ and $180^\circ$, I$_{\uparrow}$, I$_{\downarrow}$ and I$_{ph}$ all vanish, as we argued above by the PC model of Fig.~\ref{fig1}(c).

The variation of the PGE current in terms of sin(2$\theta$) or sin(2$\phi$) under linear or circular polarized light, is the typical behavior of PGE determined by the $C_s$ symmetry. According to the phenomenological theory of PGE\cite{PGE0}, whether the photocurrent is a sine or a cosine depends on the symmetry. For systems with $C_s$ symmetry, the photocurrent generated in the direction perpendicular to the mirror reflection plane is a sine, which is in perfect agreement with our first-principles results presented here. Note that the phenomenological theory has been widely adopted to explain experimental observations of LPGE and CPGE, for instances for p-GaAs/AlGaAs quantum well systems having $C_s$ symmetry\cite{PGE2}.

\begin{figure}
\centering
\includegraphics[width=0.9\columnwidth]{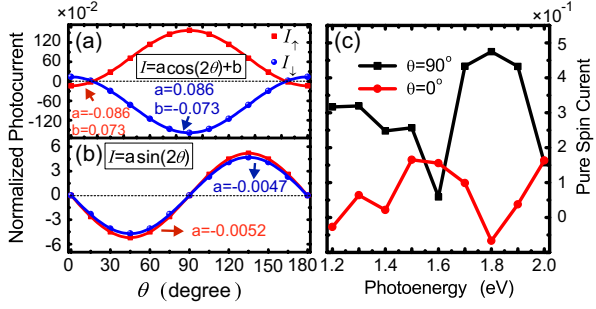}
\caption{(a) and (b) are two components of the photocurrent, proportional to cos(2$\theta$) and sin(2$\theta$), respectively, under the illumination of linearly polarized light for APC (spin-battery). Symbols are calculated data and solid lines are fitted to the data. (c) Operational points of an ideal spin-battery for both linearly and circularly polarized light. The data points give values of the photon energy and polarization angle $\theta$, where pure spin-currents without an accompanying charge-current is generated.}
\label{fig5}
\end{figure}

As argued in the qualitative discussion above, the pure spin-battery is established by the device in APC [see Fig.~\ref{fig1}(d)] for which the calculated photocurrents by LPGE and CPGE are shown in Figs.~\ref{fig4}(c,d), respectively. The total photocurrent (dark triangles) are exactly in a shape of sin(2$\theta$) for LPGE or sin(2$\phi$) for CPGE, and are well fitted by the sine function (dark lines). In contrast, for individual spin channels, I$_{\uparrow}$ and I$_{\downarrow}$ have opposite sign and appear more like a cosine function. As a result, at $\theta, \phi=0^{\circ}$, $90^{\circ}$ and $180^{\circ}$, the calculated total PGE charge-current I$_{ph}$ is essentially zero, while the spin-current I$_s$=I$_{\uparrow}-$I$_{\downarrow}$ is nonzero. Namely, the spin-battery generates a pure spin-current without an accompanying charge-current at these polarization angles. The calculated data are in perfect consistency with the qualitative discussion above.

To further understand the origin of pure spin-current generated by spin-battery, we now analyze the photocurrent which, if generated by linearly polarized light, has three components proportional to sin$^2(\theta$), cos$^2(\theta$) and sin($2\theta$), respectively [Eq.(1)]. The first two terms can be expressed in terms of cos(2$\theta$). In APC (spin-battery), the charge distribution in the device remains to have $C_s$ symmetry, since the anti-parallel spin configuration does not influence the charge distribution as mentioned above. Therefore, I$_{ph}$ should be proportional to sin($2\theta$). However, spin distribution of the system is no longer mirror-symmetric in APC, therefore the two spin-resolved currents I$_{\uparrow}$, I$_{\downarrow}$, should neither be a perfect sine function nor a perfect cosine function. For linearly polarized light, I$_{\uparrow}$ and I$_{\downarrow}$ have both cos(2$\theta$) and sin($2\theta$) components, although the former has a much larger amplitude as shown in Figs.~\ref{fig5}(a,b). It is thus concluded that I$_{\uparrow}$ and I$_{\downarrow}$ flow in opposite directions with the same magnitude at $\theta=0^{\circ}$, $90^{\circ}$ and $180^{\circ}$, as the total photocurrent I$_{ph}$=I$_{\uparrow}$+ I$_{\downarrow}$ vanishes due to its sine dependence on $2\theta$. A similar argument holds for circularly polarized light. Therefore, in APC or the spin-battery, PGE generates a pure spin-current at polarization angles $\theta, \phi=0^{\circ}, 90^{\circ}$, and $180^{\circ}$. These results are what expected from the model presented in Fig.~\ref{fig1}(d), and also understandable from the phenomenological point of view\cite{PGE0}.

Our calculation show that the linearly and circularly polarized light generates the same pure spin-current at $\theta=0^{\circ}$, also at $\theta=90^{\circ}$, which is also expected as at these angles the circularly polarized light has only one polarization vector hence equivalent to a linearly polarized light. Fig.~\ref{fig5}(c) plots the calculated operational points of the spin-battery in terms of the photon energy in the range of 1.2 eV to 2.0eV and polarization angle $\theta$. Namely, at the calculated data points, the spin-battery delivers pure spin-currents without an accompanying charge-current.

The PGE spin-battery is not limited to phosphorene but applicable in general as long as the appropriate symmetry condition is satisfied. For example, substituting the newly reported 2D puckered-arsenene\cite{as1,as2,as3} for the phosphorene should also be able to generate pure spin-current due to a similar structural symmetry. The flow of pure spin-current without an accompanying charge current is usually detected via the inverse spin-Hall effect\cite{ISH,opt4,ISH1}. For PGE spin-battery, a convenience is the characteristic sinusoidal dependence of the photocurrent which can serve as evidence for successful generation of pure spin-current, namely the PGE spin-battery not only generates pure spin-current but also provides detection by itself. Very interestingly, a recent experiment reported a persistent photocurrent in a few-layer phosphorous at zero source-drain bias\cite{BpOpt2}. While the measured data were not interpreted by the PGE physics\cite{PGE0}, the form of the observed photocurrent versus polarization angle $\theta$ suggests that it was likely due to PGE. We conclude that 2D PGE in materials without SOC such as phosphorene to be quite feasible and the PGE spin-battery realizable.

\subsection{PGE spin-battery with disorders and SOC}

\begin{figure}[!htb]
\centering
\includegraphics[width=\columnwidth]{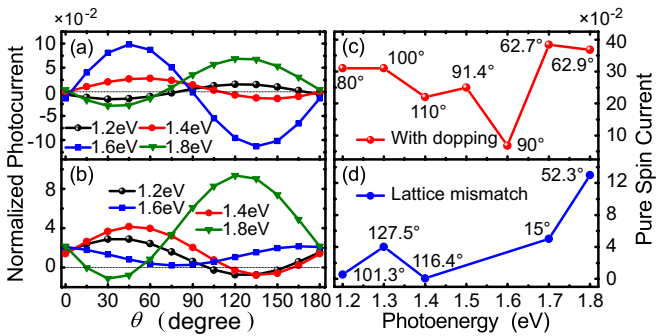}
\caption{Photocurrent at different photon energy of linearly polarized light for (a) the doping of the As atoms and (b) the displacement of the right-hand Ni(111) contact in spin-battery. (c) operational points of the non-ideal spin-battery  doped with As atom and (d)  with contact asymmetry. The data points give values of the photon energy and polarization angle $\theta$, where pure spin-currents without an accompanying charge-current is generated.}
\label{fig6}
\end{figure}

The symmetry property of the spin-battery structure is so far emphasized to interpret results. In reality, an absolutely perfect symmetry (e.g. $C_s$) is difficult to achieve due to experimental factors such as the existence of impurities and slight differences in the two Ni/phosphorene contacts. Hence the stability against such perturbations should be examined. Concerning effects of impurity, a recent experiment showed that phosphorene can be alloyed with As atoms up to 80\% As concentration\cite{as2}. We thus calculated a system by substituting an As atom for the rightmost P atom in the bottom sub-layer of the phosphorene [See Fig.~\ref{fig3}(a)]. Concerning asymmetry in the two Ni/phosphorene contacts, we calculated a system by displacing the right Ni contact and the connected Ni(111) substrate in the central region by 0.2 \AA\ along the $x$ direction, corresponding to a mismatch of 4.63\% relative to the lattice constant (4.316 \AA) of Ni(111) surface in the $x$ direction. Both these changes break $C_s$ symmetry of the original ideal spin-battery in Fig.~1 (d) of the main text. The calculated photocurrent becomes that presented in Figs.~\ref{fig6}(a,b) and several observations are in order. First,
the PGE photocurrent changes from a perfect sin(2$\theta$) of the ideal spin-battery to a form of sin(2$\theta$+$\delta \theta$)+$C$, where $\delta \theta$ and $C$ are a phase shift and a constant depending on the photon energy, as shown by different curves in Figs.~\ref{fig6}(a,b). Second, compared with results of ideal spin-battery, the photocurrent is not so much affected by the As impurity as shown in Fig.~\ref{fig6}(a), but is significantly altered due to the contact asymmetry as shown in Fig.~\ref{fig6}(b). Third and most important, although the photocurrent $I_{ph}$ is no longer a perfect sine function on the polarization angle, zero $I_{ph}$ still occurs at certain polarization angles and photon energies, as evidenced in Figs.~\ref{fig6}(a,b). In other words, the non-ideal spin-battery still generates pure spin-current even with the structural perturbations. Figs.~\ref{fig6}(c,d) present the calculated operation points of the non-ideal spin-battery in terms of the photon energy and polarization angle, this is to be compared with that of Fig.~\ref{fig4} (c) for the ideal spin-battery. We conclude that the spin-battery works whether it is ideal or perturbed, albeit at some different photon energies and polarization angles. Importantly, the signature of pure spin-current generation is when the total photocurrent $I_{ph}$ vanishes which provides the operational point of the device whether it is ideal or not.

\begin{figure}[!htb]
\centering
\includegraphics[width=0.9\columnwidth]{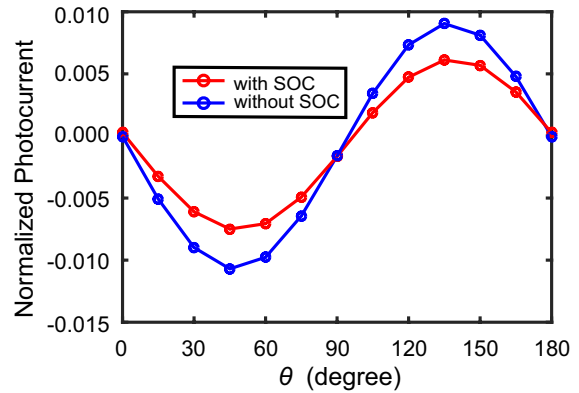}
\caption{Total photocurrent versus the polarization angle for the photon energy of 1.2 eV with and without considering the spin orbital coupling.}
\label{fig7}
\end{figure}

It is interesting to investigate if the SOC effect will significantly alter the outcome of the spin-battery. To this end, we carried out calculations for the same spin-battery system by including SOC in the first principles analysis. Results show that the photocurrent is still a perfect sin(2$\theta$), as shown in Fig.~\ref{fig7}, which qualitatively well agrees  with results where SOC is neglected.

In summary, we found that the 2D spin-battery device operated on PGE can deliver a pure spin-current without an accompanying charge-current to the outside world, which is important for 2D flexible spintronics. The device principle is implemented in the 2D material phosphorene and first principles calculations gave numerical data in excellent qualitative agreement with the PGE physics. The PGE spin-battery has excellent operational stability against structural disorder, photon energy and flux, and other material details. The 2D spin-battery is interesting as it is a device that generates pure spin-currents, an energy source that harvest photons, as well as a pure spin-current detector itself. Finally, given the recent experimental discovery of 2D ferromagnetism\cite{gong1,huang1}, the PGE spin-battery may be entirely realized by layered materials.

\vspace{0.3cm}
Y.W. is grateful to Drs. Lei Liu, Yanxia Xing and Chao-Cheng Kaun for useful discussions. This work was supported by the National Natural Science Foundation of China under grant No. 11404273 and 11774217 (Y.W.) and No. 51871156 (Y.X.), the University Grant Council (Contract No. AoE/P-04/08) of the Government of HKSAR (Y.W., J.W., H.G.).


\begin{thebibliography}{99}

\bibitem{nnano-review}
G.~Fiori, F.~Bonaccorso, G.~Iannaccone, T.~Palacios, D.~Neumaier, A.~Seabaugh,
  S.~K. Banerjee, and L.~Colombo, {Electronics based on two-dimensional
  materials}, {Nat. Nanotechnol.} ~{\bf 9}, 768 (2014).

\bibitem{flexiable}
A.~Nathan, A.~Ahnood, M.~T. Cole, S.~Lee, Y.~Suzuki, P.~Hiralal, F.~Bonaccorso,
  T.~Hasan, L.~Garcia-Gancedo, A.~Dyadyusha, S.~Haque, P.~Andrew, S.~Hofmann,
  J.~Moultrie, D.~Chu, A.~J. Flewitt, M.~J. Ferrari, A C abd~Kelly,
  J.~Robertson, G.~A.~J. Amaratunga, and W.~I. Milne, { Flexible electronics:
  The next ubiquitous platform.}, {Proc. IEEE} ~{\bf 100}, 1486 (2012).

\bibitem{spintronics}
S.~A. Wolf, D.~D. Awschalom, R.~A. Buhrman, J.~M. Daughton, S.~von Molnar,
  M.~L. Roukes, A.~Y. Chtchelkanova, and D.~M. Treger, {Spintronics: a
  spin-based electronics vision for the future}, {Science} ~{\bf 294},
1488 (2001).

\bibitem{RevSpin}
I.~Zutic, J.~Fabian, and S.~D. Sarma, { Spintronics: Fundamentals and
  applications }, {Rev. Mod. Phys.} ~{\bf 76}, 323 (2004).

\bibitem{Hu-review}
M.~Harder, Y.~Gui, and C.-M. Hu, Electrical detection of magnetization
  dynamics via spin rectification effects, {Phys. Rep.}~{\bf 661}, 1 (2016).

\bibitem{PGE0}
V.~I. Belinicher and B.~I. Sturman, {The photogalvanic effect in media
  lacking a center of symmetry}, {Sov. Phys. Usp.}~{\bf 23}, 199 (1980).

\bibitem{Agarwal}
S.~Dhara, E.~J. Mele, and R.~Agarwal,  Voltage-tunable circular photogalvanic
  effect in silicon nanowires, {Science}~{\bf349}, 726 (2015).

\bibitem{Si-Mos2}
P.~Olbrich, S.~A. Tarasenko, C.~Reitmaier, J.~Karch, D.~Plohmann, Z.~D. Kvon,
  and S.~D. Ganichev,  Observation of the orbital circular photogalvanic
  effect,   {  Phys. Rev. B}~{\bf79}, 121302 (2009).

\bibitem{Si-Mos1}
J.~Karch, S.~A. Tarasenko, E.~L. Ivchenko, J.~Kamann, P.~Olbrich, M.~Utz, Z.~D.
  Kvon, and S.~D. Ganichev,  Photoexcitation of valley-orbit currents in
  (111)-oriented silicon metal-oxide-semiconductor field-effect transistors,
  {  Phys. Rev. B}~{\bf83}, 121312 (2011).

\bibitem{PGE1}
S.~D. Ganichev, E.~L. Ivchenko, V.~V. Bel'kov, S.~A. Tarasenko, M.~Sollinger,
  D.~Weiss, W.~Wegscheider, and W.~Prettl,  {Spin-galvanic effect},   {
  Nature}~{\bf417}, 153 (2002).

\bibitem{PGE2}
S.~D. Ganichev, H.~Ketterl, W.~Prettl, E.~L. Ivchenko, and L.~E. Vorobjev,
   {Circular photogalvanic effect induced by monopolar spin orientation in
  p-GaAs/AlGaAs multiple-quantum wells},   {  Appl. Phys. Lett.}~{\bf77},
  3146 (2000).

\bibitem{PGE3}
S.~D. Ganichev, E.~L. Ivchenko, S.~N. Danilov, J.~Eroms, W.~Wegscheider,
  D.~Weiss, and W.~Prettl,  {Conversion of spin into directed electric current
  in quantum wells},   {  Phys. Rev. Lett.}~{\bf86}, 4358 (2001).

\bibitem{PGE3a}
S.~D. Ganichev and W.~Prettl,  {Spin photocurrents in quantum wells},   {
  J. Phys.: Condens. Matter}~{\bf15}, R935 (2003).

\bibitem{PGE4}
C.~Yin, H.~Yuan, X.~Wang, S.~Liu, S.~Zhang, N.~Tang, F.~Xu, Z.~Chen,
  H.~Shimotani, Y.~Iwasa, Y.~Chen, W.~Ge, and B.~Shen,  {Tunable surface
  electron spin splitting with electric double-layer transistors based on
  InN},   {  Nano Lett.}~{\bf13}, 2024 (2013).

\bibitem{PGE5}
H.~Yuan, X.~Wang, B.~Lian, H.~Zhang, X.~Fang, B.~Shen, G.~Xu, Y.~Xu, S.-C.
  Zhang, H.~Y. Hwang, and Y.~Cui,  {Generation and electric control of
  spin-valley-coupled circular photogalvanic current in WSe$_2$},   {  Nat.
  Nanotechnol.}~{\bf9},  851 (2014).

\bibitem{PGE6}
M.~Eginligil, B.~Cao, Z.~Wang, X.~Shen, C.~Cong, J.~Shang, C.~Soci, and T.~Yu,
   {Dichroic spin-valley photocurrent in monolayer molybdenum disulphide},
  {  Nat. Commun.}~{\bf6}, 7636 (2015).

\bibitem{PGE7}
C.~Kastl, C.~Karnetzky, H.~Karl, and A.~W. Holleitner,  {Ultrafast helicity
  control of surface currents in topological insulators with near-unity
  fidelity},   {  Nat. Commun.}~{\bf6}, 6617 (2015).

\bibitem{TI2}
K.~N. Okada, N.~Ogawa, R.~Yoshimi, A.~Tsukazaki, K.~S. Takahashi, M.~Kawasaki,
  and Y.~Tokura,  {Enhanced photogalvanic current in topological insulators
  via Fermi energy tuning },   {  Phys. Rev. B}~{\bf93}, 081403 (2016).

\bibitem{TI3}
J.~W. McIver, D.~Hsieh, H.~Steinberg, P.~Jarillo-Herrero, and N.~Gedik,
   {Control over topological insulator photocurrents with light polarization
  },   {Nat. Nanotechnol.}~{\bf7}, 96 (2012).

\bibitem{spc}
G.~E.~W. Bauer, E.~Saitoh, and B.~J. van Wees,  {Spin caloritronics},   {
  Nat. Mater.}~{\bf11}, 391 (2012).

\bibitem{Tian}
J.~Tian, S.~Hong, I.~Miotkowski, S.~Datta, and Y.~P. Chen,  Observation of
  current-induced, long-lived persistent spin polarization in a topological
  insulator: A rechargeable spin battery,  {  Sci. Adv.}~{\bf3}, e1602531 (2017).

\bibitem{HE1}
J.~E. Hirsch,  {Spin Hall Effect},   {  Phys. Rev. Lett.}~{\bf83},
1834 (1999).

\bibitem{opt4}
D.~Ellsworth, L.~Lu, J.~Lan, H.~Chang, P.~Li, Z.~Wang, J.~Hu, B.~Johnson,
  Y.~Bian, J.~Xiao, R.~Wu, and M.~Wu,  Photo-spin-voltaic effect,   {  Nat.
  Phys.}~{\bf12}, 861 (2016).

\bibitem{sp1}
Y.~Kajiwara, K.~Harii, S.~Takahashi, J.~Ohe, K.~Uchida, M.~Mizuguchi,
  H.~Umezawa, H.~Kawai, K.~Ando, K.~Takanashi, S.~Maekawa, and E.~Saitoh,
   {Transmission of electrical signals by spin-wave interconversion in a
  magnetic insulator},   {  Nature}~{\bf464}, 262 (2010).

\bibitem{sp2}
S.~W. Jiang, S.~Liu, P.~Wang, Z.~Z. Luan, X.~D. Tao, H.~F. Ding, and D.~Wu,
   {Exchange-dominated pure spin current transport in ${\mathrm{Alq}}_{3}$
  molecules},   {  Phys. Rev. Lett.}~{\bf115}, 086601 (2015).

\bibitem{sp3}
J.~B.~S. Mendes, O.~Alves~Santos, L.~M. Meireles, R.~G. Lacerda, L.~H.
  Vilela-Le\~ao, F.~L.~A. Machado, R.~L. Rodr\'{\i}guez-Su\'arez, A.~Azevedo,
  and S.~M. Rezende,  {Spin-current to charge-current conversion and
  magnetoresistance in a hybrid structure of graphene and yttrium iron
  garnet},   {  Phys. Rev. Lett.}~{\bf115}, 226601 (2015).

\bibitem{sps1}
K.~Uchida, S.~Takahashi, K.~Harii, J.~Ieda, W.~Koshibae, K.~Ando, S.~Maekawa,
  and E.~Saitoh,  {Observation of the spin Seebeck effect},   {  Nature},
~{\bf455}, 778 (2008).

\bibitem{sps2}
D.~Meier, T.~Kuschel, L.~Shen, A.~Gupta, T.~Kikkawa, K.~Uchida, E.~Saitoh,
  J.-M. Schmalhorst, and G.~Reiss,  {Thermally driven spin and charge currents
  in thin NiFe${}_{2}$O${}_{4}$/Pt films},   {  Phys. Rev. B}~{\bf87}, 054421 (2013).

\bibitem{sps3}
P.~Li, D.~Ellsworth, H.~Chang, P.~Janantha, D.~Richardson, F.~Shah,
  P.~Phillips, T.~Vijayasarathy, and M.~Wu,  Generation of pure spin currents
  via spin seebeck effect in self-biased hexagonal ferrite thin films,   {
  Appl. Phys. Lett.}~{\bf105}, 242412 (2014).

\bibitem{sps4}
W.~Lin, K.~Chen, S.~Zhang, and C.~L. Chien,  Enhancement of thermally injected
  spin current through an antiferromagnetic insulator,   {  Phys. Rev.
  Lett.}~{\bf116}, 186601 (2016).

\bibitem{sierra}
J.~F. Sierra, I.~Neumann, J.~Cuppens, B.~Raes, M.~V. Costache, and S.~O.
  Valenzuela,  Thermoelectric spin voltage in graphene,   {  Nat.
  Nanotechnol.}~{\bf13}, 107 (2018).

\bibitem{PRB-1}
L.~Berger,  Generation of dc voltages by a magnetic multilayer undergoing
  ferromagnetic resonance,   {  Phys. Rev. B}~{\bf59}, 11465 (1999).

\bibitem{PRB-2}
G.~Schmidt, D.~Ferrand, L.~W. Molenkamp, A.~T. Filip, and B.~J. van Wees,
   Fundamental obstacle for electrical spin injection from a ferromagnetic
  metal into a diffusive semiconductor,   {  Phys. Rev. B}~{\bf62},
  R4790 (2000).

\bibitem{PRB-3}
E.~I. Rashba,  Theory of electrical spin injection: Tunnel contacts as a
  solution of the conductivity mismatch problem,   {  Phys. Rev. B}~{\bf62},
  R16267 (2000).

\bibitem{PRB-4}
A.~Brataas, Y.~Tserkovnyak, G.~E.~W. Bauer, and B.~I. Halperin,  Spin battery
  operated by ferromagnetic resonance,   {  Phys. Rev. B}~{\bf66},
  060404(R) (2002).

\bibitem{zhang}
L.~Li, Y.~Yu, G.~J. Ye, Q.~Ge, X.~Ou, H.~Wu, D.~Feng, X.~H. Chen, and Y.~Zhang,
   {Black phosphorus field-effect transistors},   {  Nat. Nanotechnol.}
  ~{\bf9}, 372 (2014).

\bibitem{Ji}
J.~Qiao, X.~Kong, Z.-X. Hu, F.~Yang, and W.~Ji,  {High-mobility transport
  anisotropy and linear dichroism in few-layer black phosphorus},   {  Nat.
  Commun.}~{\bf5}, 4475 (2014).

\bibitem{GW}
V.~Tran, R.~Soklaski, Y.~Liang, and L.~Yang,  {Layer-controlled band gap and
  anisotropic excitons in few-layer black phosphorus},   {  Phys. Rev. B}
  ~{\bf89}, 235319 (2014).

\bibitem{BpOpt0}
M.~Buscema, D.~J. Groenendijk, G.~A. Steele, H.~S.~J. van~der Zant, and
  A.~Castellanos-Gomez,  {Photovoltaic effect in few-layer black phosphorus PN
  junctions defined by local electrostatic gating},   {  Nat. Commun.}
  ~{\bf5}, 4651 (2014).

\bibitem{BpOpt1}
M.~Buscema, D.~J. Groenendijk, S.~I. Blanter, G.~A. Steele, H.~S.~J. van~der
  Zant, and A.~Castellanos-Gomez,  {Fast and broadband photoresponse of
  few-layer black phosphorus field-effect transistors},   {  Nano Lett.}
  ~{\bf14}, 3347 (2014).

\bibitem{BpOpt2}
T.~Hong, B.~Chamlagain, W.~Lin, H.-J. Chuang, M.~Pan, Z.~Zhou, and Y.-Q. Xu,
   {Polarized photocurrent response in black phosphorus field-effect
  transistors},   {  Nanoscale}~{\bf6}, 8978 (2014).

\bibitem{BpOpt3}
Q.~Guo, A.~Pospischil, M.~Bhuiyan, H.~Jiang, H.~Tian, D.~Farmer, B.~Deng,
  C.~Li, S.-J. Han, H.~Wang, Q.~Xia, T.-P. Ma, T.~Mueller, and F.~Xia,  {Black
  phosphorus mid-infrared photodetectors with high gain},   {  Nano Lett.}
  ~{\bf16},4648 (2016).

\bibitem{small}
M.~V. Kamalakar, B.~N. Madhushankar, A.~Dankert, and S.~P. Dash,  {Low
  Schottky barrier black phosphorus field-effect devices with ferromagnetic
  tunnel contacts},   {  Small}~{\bf11}, 2209 (2015).

\bibitem{AM}
M.~Huang, M.~Wang, C.~Chen, Z.~Ma, X.~Li, J.~Han, and Y.~Wu,  {Broadband
  black-phosphorus photodetectors with high responsivity},   {  Adv. Mater.}
  ~{\bf28}, 3481 (2016).

\bibitem{vasp}
G.~Kresse and J.~Furthm\"uller,  {Efficient iterative schemes for \textit{ab
  initio} total-energy calculations using a plane-wave basis set},   {  Phys.
  Rev. B}~{\bf54}, 11169 (1996).

\bibitem{xie}
Y.~Xie, L.~Zhang, Y.~Zhu, L.~Liu, and H.~Guo,  {Photogalvanic effect in
  monolayer black phosphorus},   {  Nanotechnology}~{\bf26}, 455202 (2015).

\bibitem{Taylor}
J.~Taylor, H.~Guo, and J.~Wang,  {\textit{Ab initio} modeling of quantum
  transport properties of molecular electronic devices},   {  Phys. Rev. B}
  ~{\bf 63}, 245407 (2001).

\bibitem{acsnano-Wu}
J.~Wu, G.~K.~W. Koon, D.~Xiang, C.~Han, C.~T. Toh, E.~S. Kulkarni,
  I.~Verzhbitskiy, A.~Carvalho, A.~S. Rodin, S.~P. Koenig, G.~Eda, W.~Chen,
  A.~H.~C. Neto, and B.~Ozyilmaz,  Colossal ultraviolet photoresponsivity of
  few-layer black phosphorus,   {  ACS Nano}~{\bf9}, 8070 (2015).

\bibitem{as1}
C.~Kamal and M.~Ezawa,  {Arsenene: Two-dimensional buckled and puckered
  honeycomb arsenic systems},   {  Phys. Rev. B}~{\bf91}, 085423 (2015).

\bibitem{as2}
B.~Liu, M.~Kopf, A.~N. Abbas, X.~Wang, Q.~Guo, Y.~Jia, F.~Xia, R.~Weihrich,
  F.~Bachhuber, F.~Pielnhofer, H.~Wang, R.~Dhall, S.~B. Cronin, M.~Ge, X.~Fang,
  T.~Nilges, and C.~Zhou,  {Black arsenic phosphorus: layered anisotropic
  infrared semiconductors with highly tunable compositions and properties},
  {  Adv. Mater.}~{\bf27}, 4423 (2015).

\bibitem{as3}
Z.~Zhu, J.~Guan, and D.~Tom\'{a}nek,  {Structural transition in layered
  As$_{1-x}$P$_x$ compounds: a computational study},   {  Nano Lett.}
  ~{\bf15}, 6042 (2015).

\bibitem{ISH}
S.~O. Valenzuela and M.~Tinkham,  {Direct electronic measurement of the spin
  Hall effect },   {  Nature}~{\bf442}, 176 (2006).

\bibitem{ISH1}
D.~Sun, J.~v. Schootech, M.~Kavand, H.~Malissa, C.~Zhang, M.~Groesbeck,
  C.~Boehme, and Z.~V. Vardeny,  { Inverse spin Hall effect from pulsed spin
  current in organic semiconductors with tunable spin-orbit coupling},   {
  Nat. Mater.}~{\bf15}, 863 (2016).

\bibitem{gong1}
C.~Gong, L.~Li, Z.~Li, H.~Ji, A.~Stern, Y.~Xia, T.~Cao, W.~Bao, C.~Wang,
  Y.~Wang, Z.~Q. Qiu, R.~J. Cava, S.~G. Louie, J.~Xia, and X.~Zhang,
   Discovery of intrinsic ferromagnetism in two-dimensional van der waals
  crystals,   {  Nature}~{\bf546}, 265 (2017).

\bibitem{huang1}
B.~Huang, G.~Clark, E.~Navarro-Moratalla, D.~R. Klein, R.~Cheng, K.~L. Seyler,
  D.~Zhong, E.~Schmidgall, M.~A. McGuire, D.~H. Cobden, W.~Yao, D.~Xiao,
  P.~Jarillo-Herrero, and X.~Xu,  Layer-dependent ferromagnetism in a van der
  waals crystal down to the monolayer limit,   {  Nature}~{\bf546}, 270 (2017).

\end{thebibliography}

\end{document}